\documentclass[12pt]{article}
\usepackage{amssymb}
\def\thefootnote{\fnsymbol{footnote}}
\newcommand {\ee}{\end{equation}}
\newcommand {\bea}{\begin{eqnarray}}
\newcommand {\eea}{\end{eqnarray}}
\newcommand {\nn}{\nonumber \\}

\newcommand {\tr}{{\rm tr\,}}

\newcommand {\pl}{\partial}

\newcommand {\vp}{\varphi}

\newcommand {\al}{\alpha}
\newcommand {\be}{\beta}
\newcommand {\ga}{\gamma}

\newcommand {\la}{\lambda}

\newcommand {\si}{\sigma}

\newcommand {\del}  {\delta}

\newcommand {\half}{ {\frac{1}{2}} }

\newcommand {\Lcal}{{\cal L}}

\newcommand {\Ncal}{{\cal N}}

\def\overleftarrow#1{\vbox{\ialign{##\crcr
 $\leftarrow$\crcr\noalign{\kern-1pt\nointerlineskip}
 $\hfil\displaystyle{#1}\hfil$\crcr}}}

\newcommand {\labar}{{\bar \lambda}}

\newcommand {\sibar}{{\bar \sigma}}
\newcommand {\xibar}{{\bar \xi}}

\newcommand {\alp}{{\alpha'}}
\newcommand {\bep}{{\beta'}}
\newcommand {\gap}{{\gamma'}}
\newcommand {\q}    {\quad}
\newcommand {\qq}   {\quad\quad}
\newcommand {\qqq}   {\quad\quad\quad}

\usepackage{graphicx}
\usepackage{latexsym}
\begin{document}
\begin{flushright}
%***  Ver.05.01.16  ***
June 2005
%hep-th/0409193 \\
%US-02-05
\end{flushright}

\vspace{0.5cm}

\begin{center}

{\Large\bf 
%SUSY Effective Potential and Radius Stabilization in Warped Extra Dimensions}
A New Mechanism for Radius Stabilization in\\ Warped Supersymmetry}

\vspace{1.5cm}
%{\large Note by S.I.}
{\large Shoichi ICHINOSE
         \footnote{
E-mail address:\ ichinose@u-shizuoka-ken.ac.jp
                  }
}\ and\ 
{\large Akihiro MURAYAMA$^\ddag$
         \footnote{
E-mail address:\ edamura@ipc.shizuoka.ac.jp
                  }
}
\vspace{1cm}

{\large 
Laboratory of Physics, 
School of Food and Nutritional Sciences, 
University of Shizuoka, 
Yada 52-1, Shizuoka 422-8526, Japan
 }

$\mbox{}^\ddag${\large
Department of Physics, Faculty of Education, Shizuoka University,
Shizuoka 422-8529, Japan
}
\end{center}

\vspace{2cm}

\begin{center}
{\large Abstract}
\end{center}

A new mechanism for the radius stabilization is proposed in a 5D super-Yang-Mills model in the warped background of
$\mbox{AdS}_5$. Dominant 1-loop contribution to the supersymmetric 4D effective potential is estimated to depend on
the radion. The minimization of the effective potential incorporating the tree potential and a Fayet-Iliopoulos $D$-term
for the gauge group U(1) reveals an interesting case that the radius is stabilized at a length corresponding to an
intermediate mass scale $10^{11-13}$ GeV.

\newpage
\renewcommand{\thefootnote}{\arabic{footnote}}
\setcounter{footnote}{0}

\noindent {{\bf 1}\q{\bf Introduction}}\\

Supersymmetry (SUSY) and extra dimensions are attractive extensions of particle theories beyond the Standard Model
(SM). They are expected to solve the gauge hierarchy problem and pave the way to the unified theory involving gravity. A
bulk-boundary system has been extensively studied in this context. Many attempts in this direction
involve the compactification radius and it is necessary to find the mechanism for the radius to be stabilized. One of
interesting possibilities is to introduce scalar fields (or hypermultiplets for the case of SUSY) in the bulk and minimize
an induced radion potential \cite{{GW},{AHSW},{LuSun0},{LuSun1},{GoLuNg},{MarOka},{EMS}}. Another important issue of
the extra dimension is the problem of power law (linear) running of gauge couplings. It is well-known that in
5-dimensional (5D) Yang-Mills (YM) theories with mirror-plane boundaries the linear running of gauge coupling constants
takes place when the compact extra dimension is flat \cite{{Tay-Ven},{Dienes}}, while the running remains logarithmic
\cite{{Pom},{RSchw}} in a model of $\mbox{AdS}_5$ with the warped extra dimension of Randall-Sundrum (RS)
\cite{RS}. The origin of the power law is a "democratic" summation of Kaluza-Klein (KK) excitation modes of bulk
propagators in loop diagrams. In $\mbox{AdS}_5$, on the contrary, the ultraviolet (UV) cutoff for the 4D momentum
integral depends on a position in the fifth dimension, which restricts the effective range of KK modes and suppresses
the growth of gauge couplings. 

Recently, we have made some investigations into the bulk-boundary system \cite{{Ichi-PRD65}, {Ichi-PRD66},
{Ichi-Mura-PL1}, {Ichi-Mura-PL2}, {Ichi-Mura-PL3}, {Ichi-Mura-NP}}. Among them, in \cite{Ichi-Mura-PL2} a 1-loop 4D
effective potential has been estimated in the framework of 5D super-YM model compactified on an orbifold
$S^1/Z_2$ of size $l$ by Mirabelli-Peskin \cite{Mirab} and Hebecker \cite{Heb} using auxiliary field tadpole method
(AFTM) \cite{Miller}. The resultant effective potential has the following characteristic features: i) Its dominant part 
diverges not logarithmically but linearly, that is, it is proportional to an UV cutoff scale ${\mit\Lambda}$. ii) It satisfies
a SUSY boundary condition which means that it vanishes when the auxiliary fields have zero vacuum expectation values
(VEVs) making SUSY restored. iii) It is also proportional to $l$ and goes to zero as $l \rightarrow 0$ for it is obtained in
AFTM by reducing the 5D tadpole amplitudes to 4D. The linear divergence of i) is considered to have the same origin with
the linearity of running of gauge couplings in the flat extra dimension.

The appearance of linear divergence requires a careful treatment. Its field theoretical interpretations have been given
in \cite{Ichi-Mura-NP}. Namely, although the 5D Mirabelli-Peskin model \cite{Mirab} is perturbatively
unrenormalizable, the renormalization works well in the limit $l \rightarrow 0$ as far as the 4D world is concerned. 
For nonzero $l$, the UV cutoff of the 4D momentum can essentially be given by the inverse of $l$ provided
$l{\mit\Lambda} \lesssim 1$. As we have $l{\mit\Lambda} \gg 1$ in \cite{Ichi-Mura-PL2}, we should claim the model to
be an effective theory valid up to the scale ${\mit\Lambda}$ such that an ordinary 4D super-YM model is restored for
$l \rightarrow 0$. For $M_W \ll l^{-1} \ll {\mit\Lambda}$, the linear divergence is greatly suppressed to generate a
finite 1-loop contribution which is nothing but a quantum effect of the bulk to the boundary. 

Another interesting observation in \cite{Ichi-Mura-PL2} is that if a Fayet-Iliopoulos (FI) $D$-term is introduced as a
probe for examining the physical property of the effective potential for the gauge group U(1), there is a case such
that the potential is minimized at a specific value of $l$ where the potential vanishes. Thus, we have a true vacuum
where SUSY is restored for a nonzero value of $l$. In fact, this $l$ corresponds to an intermediate mass scale $\approx
10^{11}$ or $10^{14}$ GeV for the cutoff mass scale ${\mit\Lambda}= M_{GUT}$ or $M_{Pl}$, respectively. It should be
noteworthy that whether the FI-term violates SUSY or not is controlled by a sign of the $D$-term multiplied by the
gauge coupling. 

Therefore, it is worthwhile to investigate whether the effective potential grows linearly, as increasing
${\mit \Lambda}$ in $\mbox{AdS}_5$, or not. It is also interesting to know how the phenomenon of SUSY restoration at
the specific $l$ is connected with the radius stabilization. 

In this paper, we propose a simple model which implements a new mechanism for the radius stabilization by making use
of the linear growth of 1-loop effective potential $V_{1-loop}$. We will be based on the 5D super-YM model with
the warped extra dimension \cite{RS}. The field content is the same as that of \cite{Mirab}, i.e., we have 4D chiral
matter multiplets on one brane in addition to the 5D gauge multiplet in the bulk. It should be stressed that we do {\it not}
have any hypermultiplet in the bulk. The result will show a possibility to give a linear cutoff dependence as in the flat
extra dimension and to realize the radius stabilization for the gauge group U(1). This is explained as follows: The main
contribution to our effective potential is from the bulk propagator of the scalar component $\mit\Phi$ of 5D gauge
supermultiplet and the interaction of $\mit\Phi$ with the boundary matter field $\phi$'s takes place only in
the 4D boundary. Therefore, the relevant loop amplitude including the $\mit\Phi$-propagator can incorporate the whole
KK modes up to the UV cutoff ${\mit\Lambda}$ without being restricted by the extra dimension-dependent cutoff
\cite{RSchw} in the RS \cite{RS} background when the matter superfield is placed on the so-called "Planck brane". The
fields on the Planck brane do not directly couple with the radion \cite{LuSun1}. However, as the 4D potential
$V_{1-loop}$ is derived from 5D and subject to the bulk quantum effects, it does depend on $l; V_{1-loop}\propto
l{\mit\Lambda}$, where $l$ is related with the VEV of the radion.

In our model, we will not consider the full 5D supergravity (SUGRA) but only a part of it, i.e., the radion superfield $T$.
Therefore, our formulation should be regarded as that of rigid SUSY on nontrivial gravitational backgrounds. In
particular, we assume that the backreaction of fields in the bulk-boundary system to the background metric can
be neglected.
\\

\noindent{{\bf 2}\q {\bf 5D super-Yang-Mills model in $\mbox{AdS}_5$}}\\

We will consider the 5D space-time with the signature $(+----)$.  The extra dimension $x^5$ is compactified on an
orbifold $S^1/Z_2$ of size $l$. We take an anti-deSitter space metric as in \cite{RSchw}:
\begin{eqnarray}
ds^2=\frac{1}{k^2z^2}(\eta_{\mu\nu} dx^\mu dx^\nu - dz^2), \label{AdS01}
\end{eqnarray}
where $z \equiv \exp(k\vert x^5\vert)/k=\exp(kr\vert y\vert)/k \ (y\equiv (\pi/l)x^5, -\pi \leq y \leq \pi)$ with $r$
being the radius of
$S^1$ which we identify with a real part of scalar component of radion chiral  multiplet $T$
\cite{{LuSun0},{LuSun1},{MP}}:
\begin{eqnarray}
T=r+ib+\theta\psi_R+\theta^2 F_T,
\end{eqnarray}
where $b, \psi_R$ and $F_T$ denote the fifth component of graviphoton, the fifth component of right-handed gravitino
and a complex auxiliary field, respectively.  We regard $l/\pi$ as the VEV of $r$, i.e.,
$l\equiv\pi\langle r\rangle$. We have two branes: a "Planck brane" at $z=1/k \ (y=0)$ and a "TeV brane" at $z=1/q \
(y=\pi)$. $q$ is related to  $l$ by
\begin{eqnarray}
q= k\exp (-kl), \label{AdS00}
\end{eqnarray} 
and $1/k$ is the AdS curvature radius. 

Using the superconformal method of SUGRA \cite{{FER}, {STELL}, {CREM-JULIA}, {CREM}, {KUGO}}, the
most general lagrangian describing 4D SUGRA coupled to $T$ can be written as
\begin{eqnarray}
{\cal L}_{4D SUGRA} = \int d^4\theta\vp^\dag\vp f(T^\dag, T), 
\end{eqnarray} 
where
\begin{eqnarray}
\vp = 1 + \theta^2 F_\vp,
\end{eqnarray}
is the conformal compensator. Expanding ${\cal L}_{4D SUGRA}$ in component fields as
\begin{eqnarray}
{\cal L}_{4D SUGRA} = \sqrt{-g}\left[-\frac{1}{6}fR(g) + \cdots\right],
\end{eqnarray}
where $g = \det (g_{\mu\nu})$ is the determinant of 4D metric\footnote{Here we have the freedom to make a Weyl
rescaling of the metric but it has been shown in \cite{LuSun1}to be identified with the 4D metric corresponding to
(\ref{AdS01}).} and $R(g)$ is the 4D Ricci scalar, and matching it to the 4D effective  lagrangian
%\begin{eqnarray}
%{\cal L}^{eff}_{4D} = - \frac{M_5^3}{k}\sqrt{-g}\left[\left(1- e^{-2\pi kr(x)}\right)R(g) + \cdots\right],
%\end{eqnarray} 
reduced from 5D SUGRA lagrangian 
\begin{eqnarray}
{\cal L}_{5D SUGRA} = - M_5^3\sqrt{G}[\half R(G) + \cdots],
\end{eqnarray} 
by integrating over $y$, we obtain
\begin{eqnarray}
f = \frac{3M_5^3}{k}\left(1- e^{-\pi k(T+T^\dag)}\right)= \frac{3M_5^3}{k}\left(1- e^{-2\pi kr}\right),
\end{eqnarray}
where $M_5,G = \det (G_{MN})$ and $R(G)$ denote the 5D Planck scale, the determinant of 5D metric and the 5D Ricci
scalar, respectively. The VEV of $f$ is related to the effective 4D Planck scale:
\begin{eqnarray}
M_{Pl}^2 = \langle f(r) \rangle = \frac{3M_5^3}{k}\left(1- e^{-2\pi k\langle r\rangle}\right) =  
\frac{3M_5^3}{k}\left(1- \frac{q^2}{k^2}\right). \label{AdS22}
\end{eqnarray} 
The $T$ and $T^\dag$-derivative of $f(r), \pl^2f/\pl T^\dag\pl T$, serves a coefficient of the radion kinetic term in the
4D SUGRA.

It is assumed that $q \ll k\lesssim M_5 \lesssim M_{Pl}$ and $kl \approx 35$ in order for
$q \leq $O(TeV) to be realized. %Therefore, $l^{-1}$ is only one order or so smaller than $M_{Pl}$.

We put a 5D gauge supermultiplet in the bulk which is made of a vector field $A^M\ (M=0,1,2,3,5)$,  a scalar
field $\mit\Phi$, a doublet of symplectic Majorana fields $\la^i\ (i=1,2)$, and a triplet of auxiliary scalar fields $X^a\
(a=1,2,3)$. All bulk fields are of the adjoint representation of the gauge group: $A^M = A^{M\al}T^\al$, etc.,
$\tr[T^\al T^\be]=\del^{\al\be}/2$ and ${\cal D}_M\mit\Phi = \pl_M\mit\Phi - ig[A_M,\mit\Phi]$. We project  out
$\Ncal=1$ SUSY multiplets by assigning $Z_2$-parity to all fields in accordance with the 5D SUSY. A consistent choice
is given as: $P=+1$ for $A_\mu (\mu=0,1,2,3), \la^1_L, X^3$;  $P=-1$ for $A_5, {\mit\Phi}, \la^2_L, X^1, X^2$.
(The fields of $P=-1$ vanish on the branes $x^5 = 0, l$.) Then, $V\equiv(A_\mu,\la^1_L,(2l)^{-\half}D) \ (D\equiv
(2l)^\half( X^3-{\cal D}_5{\mit\Phi}))$ and
${\mit\Sigma}\equiv({\mit\Phi} + iA_5,-i\sqrt{2}\la^2_R,X^1 + iX^2)$ constitute an $\Ncal =1$ vector supermultiplet
in Wess-Zumino gauge and a chiral scalar supermultiplet, respectively. We do not have any hypermultiplet in the bulk
which is usually used in order for the radius stabilization to be realized. Our model is simpler in this respect and the
radius stability will be discussed without hypermultiplets.  

We introduce on one of two branes a 4D massless chiral supermultiplet $S \equiv (\phi,\psi,F)$ of the fundamental
representation where $\phi,\psi$ and $F$ stand for a complex scalar field, a Weyl spinor and an auxiliary field of
complex scalar, respectively. 

The 5D SUSY action which describes the couplings of the radion with the bulk and brane fields can be derived by using the
superconformal approach \cite{{LuSun0},{LuSun1},{MP},{CORR},{ABE}}. We follow the actions given by Marti-Pomarol
\cite{MP}. The gauge invariant actions for 5D gauge supermultiplet and 4D chiral matter supermultiplet are then given by
\begin{eqnarray}
S_{gauge} & = & \int d^4xdz\ \tr\left[\left\{\int d^2\theta \half TW^\al W_\al + \mbox{h.c.}\right\}\right.\nn
 &&\qqq\qqq + \left. \int d^4\theta \frac{(kz)^{-2}}{T+T^\dag}\left(\pl_y V + \frac{1}{\sqrt{2}}({\mit
\Sigma} + {\mit \Sigma}^\dag)\right)^2 \right], \label{AdS33}
\end{eqnarray} 
and
\begin{eqnarray}
S_{matter} & = & S_{matter}^{(1/k)} \q {\mbox or} \q S_{matter}^{(1/q)},\\
S_{matter}^{(1/k)} & = & \int d^4xdz \left[\int d^4\theta (S^\dag e^{gV}S+2\xi V) \right.\nn
&& \qqq\qqq\qqq + \left.\left\{\int d^2\theta W(S) + \mbox{h.c.}\right\}\right]\del(z-\frac{1}{k}), \\
S_{matter}^{(1/q)} & = & \int d^4xdz \left[\int d^4\theta e^{-\pi k(T+T^\dag)}(S^\dag
e^{gV}S+2\xi V) \right. \nn && \qqq\qqq\qqq + \left.\left\{\int d^2\theta e^{-3\pi kT}W(S) +
\mbox{h.c.}\right\}\right]\del(z-\frac{1}{q}),
\end{eqnarray}
respectively, where $W_\al=(-1/4){\bar D}^2D_\al V,\ W(S)$ is a superpotential and $\xi \neq 0$ only in case of the
gauge group containing U(1) giving rise the FI $D$-term. 

After the relevant rescaling of the fields in (\ref{AdS33}), we obtain
\begin{eqnarray}
S_{gauge} & = & \int d^4xdz \sqrt{G}\ \tr\left[ - {\textstyle \half}({F_{MN}})^2+ ({\cal D}_M{\mit\Phi})^2
+ (i{\labar}^i\gamma^M \nabla_M\lambda^i)  +  (X^a)^2\right. \nn
&& \qqq\qqq\qq \left.+ (g{\labar}^i [{\mit\Phi},\la^i])- 2m_{\mit\Phi}^2{\mit\Phi}^2
-2im_\la{\labar}^i(\sigma^3)^{ij}\la^j\right], \label{AdS03}
\end{eqnarray}
where ${\cal D}_M \equiv \pl_M - igA_M$ (with appropriate adjoint action of $A_M$ being implied), $\nabla_M  \equiv
{\cal D}_M + {\mit\Gamma}_M$ with ${\mit\Gamma}_M$ being the spin connection \cite{{Shuster},{Gher-Pom}},
$m_{\mit\Phi}^2 = -4k^2 + 4k\{\del(z - 1/k)-\del(z - 1/q)\}$ and $m_\la = (1/2)k\epsilon(z-1/k)$. The action 
(\ref{AdS03}) has no direct coupling of radion to the gauge supermultiplet for the sake of the rescaling.

The radion does not directly couple to the matter fields in $S_{matter}^{(1/k)}$, too, since the warp factor
is unity at $z=1/k$.  Expanding $S_{matter}^{(1/k)}$ into component fields, we obtain
\begin{eqnarray}
S_{matter}^{(1/k)} & = & \int d^4xdz \sqrt{G}[{\cal D}_\mu\phi^\dag {\cal D}^\mu\phi+\psi^\dag i\sibar^\mu 
{\cal D}_\mu \psi+F^\dag F \!-\!\sqrt{2}g(\phi^\dag\la^T_L\si^2\psi+\psi^\dag \si^2\la^*_L\phi) \nn
&&\!\!\! + g\phi^\dag D\phi +\xi D + \{{\textstyle\half}\lambda_{\alp\bep\gap}(\phi_\alp\phi_\bep F_\gap -
\psi_\alp\psi_\bep\phi_\gap) +\mbox{h.c.}\}]\del(z - 1/k), \label{AdS04}
\end{eqnarray}
where we have taken the following superpotential:
\begin{equation}
W(S) = \frac{\lambda_{\alp\bep\gap}}{3!}S_\alp S_\bep S_\gap,\label{AdS05}
\end{equation}
with the coefficient $\lambda_{\alp\bep\gap}$ being such that the gauge symmetry is respected.\footnote{The sign of
$\lambda$ here is taken to be opposite to that in \cite{Ichi-Mura-PL2}.} The expansion of $S_{matter}^{(1/q)}$ is the
same as (\ref{AdS04}) except the effect of warp factors.

$S_{gauge}$ is invariant under the 5D SUSY transformations \cite{{Mirab},{Heb},{Shuster}}
\begin{eqnarray}
\del A^M & = & i\xibar^i\ga^M\la^i, \nn
\del{\mit\Phi} & = & i\xibar^i\la^i, \nn
\del\la^i & = & (\sigma^{MN}F_{MN}-\ga^M{\cal D}_M{\mit\Phi})\xi^i-i(X^a\sigma^a)^{ij}\xi^j +
4im_\la{\mit\Phi}(\sigma^3)^{ij}\xi^j, \nn
\del X^a & = & \xibar^i(\sigma^a)^{ij}\ga^M{\cal D}_M\la^j - i[{\mit\Phi}, \xibar^i(\sigma^a)^{ij}\la^j], \label{AdS20}
\end{eqnarray}
where $\sigma^{MN} \equiv \frac{1}{4}[\ga^M,\ga^N]$. On the other hand, $S_{matter}$ is invariant under the
standard 4D SUSY transformations \cite{{Mirab},{Heb}} and the SUSY transformations of bulk fields (\ref{AdS20})
reduced to 4D. 

In subsequent sections, we will develop an effective theory in a sense mentioned in Section 1 on the basis of the above
actions. 
\\

\noindent{{\bf 3}\q {\bf Estimation of 1-loop diagrams}}\\

In computing the 1-loop effective potential $V_{1-loop}$ in \cite{Ichi-Mura-PL2}, the method AFTM \cite{Miller} has
been used. When there arises no quadratic divergence, the main contribution to  $V_{1-loop}$ has proved to
come from the diagrams in Fig.1. There the bulk propagator of ${\mit\Phi}$ contains the KK modes and gives rise to the
linear divergence in the {\it flat} background of the extra dimension. Since it would be allowed to use AFTM \cite{Miller}
in the {\it warped} background, too, we will estimate here the diagrams in Fig.1 which are thought to contribute
dominantly to
$V_{1-loop}$. 

Following \cite{RSchw}, we represent the propagator in the momentum space for 4D part, but in the position space for
the part of the fifth dimension. 
\begin{figure}
\begin{center}
\includegraphics[width=150.mm,height=65.55mm]{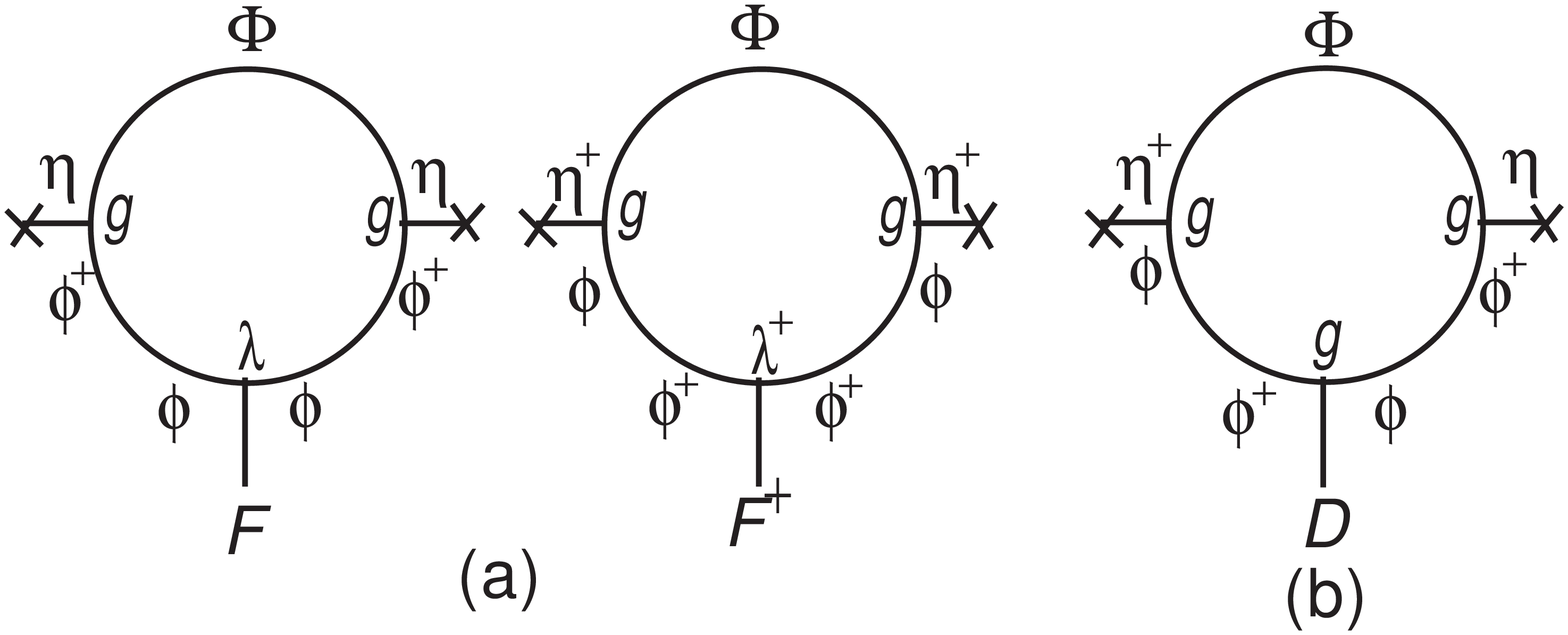}
%\BoxedEPSF{RadStazu1.eps scaled 400}
%\centerpicture 130.5mm by 88.55mm ( Zu001 scaled 1000)
\end{center}
\caption{1-loop diagrams dominantly contributing to $V_{eff}$, (a) $F$- and $F^\dag$-tadpoles and (b) $D$-tadpole.
The mark {\sf x} indicates VEVs $\eta$ or $\eta^\dag$.}
\label{Fig1}
\end{figure}
The ${\mit\Phi}$-propagator is then expressed by the Green's function $G_p(z, z')$ which satisfies
\begin{eqnarray}
\left[\pl_z^2 - \frac{3}{z}\pl_z + p^2 - \frac{m_{\mit\Phi}^2}{(kz)^2}\right]G_p(z, z') = (kz)^3\del(z-z'),
\label{AdS051}
\end{eqnarray}
where $p_\mu = i\pl_\mu$ is the four-momentum. As $m_{\mit\Phi}^2=-4k^2$ for $\Lcal_{blk}$ (\ref{AdS03}) to be
SUSY, the homogeneous solutions are
\begin{eqnarray}
G_p(u,v) = u(AK_0(pu)+BI_0(pu))=v(CK_0(pv)+DI_0(pv)), \label{AdS006}
\end{eqnarray}
where $u \equiv \mbox {min}(z,z')$, $v \equiv \mbox {max}(z,z')$ and a Wick rotation has been performed. With the
$Z_2$-parity of ${\mit\Phi}$ being odd, the solutions must obey Dirichlet boundary conditions:
\begin{eqnarray}
G_p(\frac{1}{k},v) = G_p(u, \frac{1}{q}) = 0. \label{AdS07}
\end{eqnarray}
Then, by matching the two solutions over the delta function, we obtain
\begin{eqnarray}
G_p(u,v) = k^3u^2v^2\frac{\{AK_0(pu)+BI_0(pu)\}\{CK_0(pv)+DI_0(pv)\}}{AD+BC}, \label{AdS06}
\end{eqnarray}
where
\begin{eqnarray}
A = I_0\left(\frac{p}{k}\right),\ B = - K_0\left(\frac{p}{k}\right),\ C = I_0\left(\frac{p}{q}\right),\ D = -
K_0\left(\frac{p}{q}\right).
\label{AdS016}
\end{eqnarray}

Making use of the Feynmann rules and regulating the loop integrals following \cite{RSchw}, the contribution of Fig.1 to
the 4D effective potential is written as
\begin{eqnarray}
&&V_{eff}(\mbox{Fig.1}) = {\cal A}^{(w)}\frac{l}{2}\int_\mu^{\mit\Lambda}\frac{d^4p}{(2\pi)^4}
\frac{1}{p^4} I(p,1/k,{\mit\Lambda}/(kp), w), \label{AdS08}\\
&&\q I(p, a,b,w)  \equiv \int_a^b\frac{du}{(ku)^6}\int_u^b\frac{dv}{(kv)^6}
\del(u-w)\del(v-w)k^2uv\frac{\pl^2G_p(u,v)}{\pl u\pl v}, \label{AdS09}
\end{eqnarray}
where ${\mit\Lambda}$ denotes the UV cutoff. We have assumed $\mu(\mbox{infrared cutoff})={\mit\Lambda}q/k$ and
${\mit\Lambda}\gtrsim k$. The factor $l$ in (\ref{AdS08}) is to reduce the effective potential from 5D to 4D.
The factor ${\cal A}^{(w)}$ is defined for $w=1/k$, i.e., for the matter fields placed on the "Planck brane", as
\begin{eqnarray}
{\cal A}^{(1/k)} \equiv \half g^2\{(T^\al\eta)^t \lambda f T^\al\eta + \eta^\dag T^\al \lambda^\dag
f^\dag (\eta^\dag T^\al)^t\} - g^3(\eta^\dag T^\al T^\be T^\al\eta)d_\be,  
\end{eqnarray}
where $\eta \equiv \langle \phi \rangle, \ f \equiv \langle F \rangle$ and $d \equiv \langle D \rangle$ are VEVs,
$g/\sqrt{2l}$ has been rewritten as dimensionless $g, \ (\lambda f)_{\alp\bep}\equiv \lambda_{\alp\bep\gap}f_\gap$
and $(\lambda^\dag f^\dag)_{\alp\bep}\equiv \lambda^\dag_{\alp\bep\gap}f^\dag_\gap$. For $1/q$, i.e., for the matter
fields placed on the "TeV brane", ${\cal A}^{(1/q)}$ is defined similarly as ${\cal A}^{(1/k)}$ except the effect of the
warp factors, which need not be specified here since the matter fields on the "TeV brane" do not contribute to the 1-loop
effective potential. Namely, (\ref{AdS09}) can remain nonzero only when $p = \mu$ which is nothing but the lower bound
of the integral.   

If the matter fields are placed on the "Planck brane", i.e., $w = 1/k$, we split the integral into three regions; $p \lesssim
ck, ck \lesssim p \lesssim c'k$ and $c'k \lesssim p$ with $0 < c \ll 1 \ll c'$. Then, for the region $p \lesssim ck$,
\begin{eqnarray}
\int_\mu^{ck}\frac{d^4p}{(2\pi)^4}\frac{1}{p^4}I(p,1/k,c/p,1/k) &\approx&
\int_\mu^{ck}\frac{d^4p}{(2\pi)^4}\frac{1}{p^4} \left\{2p + \frac{k}{\log (2k/p)-\ga}\right\} \nn
& \approx
&\frac{k}{4\pi^2}\log\left\{\log\left(\frac{\mu}{2k}\right)\mbox{\Large{/}}\log\left(\frac{c}{2}\right)\right\},
\label{AdS11}
\end{eqnarray}
while for $c'k \lesssim p$,
\begin{eqnarray}
&&\int_{c'k}^{\mit\Lambda}\frac{d^4p}{(2\pi)^4}\frac{1}{p^4}I(q,1/k,{\mit\Lambda}/(kp),1/k) \nn
&\approx&\int_{c'k}^{\mit\Lambda}\frac{d^4p}{(2\pi)^4}\frac{1}{p^4}\left\{2k +
p\coth\left(\frac{p}{q}\right)\right\}\nn
& = & \frac{1}{16\pi^2}\left\{2k\log\left(\frac{{\mit\Lambda}}{c'k}\right) +
{\mit\Lambda} - q\log\left(2\sinh\left(\frac{c'k}{q}\right)\right)\right\}, \label{AdS12}
\end{eqnarray}
where we have used the following limits of the modified Bessel functions: $I_n(z) \rightarrow (z/2)^n, \ K_{n\neq 0}(z)
\rightarrow (z/2)^{-n/2}, \ K_0(z) \rightarrow -(\ga + \log(z/2))$ for $z \rightarrow 0$ and $I_n(z) \rightarrow
\exp(z)/\sqrt{2\pi z}, \ K_n(z) \rightarrow \sqrt{\pi/2z}\exp(-z)$ for $z \rightarrow \infty.$

Thus, we obtain for $w = 1/k$
\begin{eqnarray}
V_{eff}(\mbox{Fig.1}) = \frac{{\cal A}}{32\pi^2}\left\{2lk\log\left(\frac{{\mit\Lambda}}{c'k}\right) 
+l{\mit\Lambda}\right\} + \mbox{finite terms}. \label{AdS13}
\end{eqnarray} 
It is remarkable that a term proportional to $l{\mit\Lambda}$ in the effective potential does appear when the matter
supermultiplet lives on the "Planck brane". This is because the whole KK modes up to ${\mit\Lambda}$ has been {\it
democratically} summed in the loop amplitude without being restricted by the extra dimension-dependent cutoff
\cite{RSchw}. It should be stressed that this is due to quantum effects of the bulk. For $l \rightarrow 0$, we
have $V_{eff}(\mbox{Fig.1}) \rightarrow 0$ so that an ordinary 4D super-YM  model is restored leaving only the
contribution from tadpoles confined on the brane behind. Such 4D tadpoles present the usual logarithmic
divergence\footnote{It is the term proportional to $\tr {\cal G}_+$ in \cite{Ichi-Mura-PL2}.} in addition to the
quadratic divergence which arises for the gauge group with U(1) factor and will spoil the nonrenormalization theorem.
\\

\noindent{{\bf 4}\q {\bf Minimization of the effective potential}}\\

Hereafter, we choose the gauge group to be U(1). We include additional chiral supermultiplets with similar actions as
(\ref{AdS04}), arrange their U(1)-charges and VEVs such that the quadratic divergence does not arise. This is
easily embodied by a toy model in which three chiral superfields $S_1, S_2, S_3$ on the "Planck brane" with
U(1)-charges $g/2, g/2, -g$ and with the superpotential
\begin{eqnarray}
W = \frac{1}{3}\left(\frac{\lambda}{3!}S_1 S_2 S_3 + \frac{\lambda'}{3}S^2_1 S_3 +
\frac{\lambda''}{3}S^2_2 S_3\right),\label{AdS005}
\end{eqnarray}
have VEVs $\langle \phi_1\rangle = \langle \phi_2\rangle = \eta, \langle \phi_3\rangle = 0, \langle F_1\rangle =
\langle F_2\rangle = 0$ and $\langle F_3\rangle = f$. In the following, we take this toy model
assuming $\lambda = \lambda' = \lambda''$ for simplicity. Then, the  effective potential to be minimized reads
\cite{Ichi-Mura-PL2}
\begin{eqnarray}
V_{eff} = && \!\!\!\!\!\!\!\!\!\! V_{tree} + V_{1-loop} + V_{FI}, \label{AdS14}\\
V_{tree} & = & - f^\dag f - \half\lambda f\eta^2 -
\half\lambda^\dag f^\dag\eta^{\dag 2} - \half d^2 - g\eta^\dag d\eta, \label{AdS15}\\
V_{1-loop} & = &  \al \{(\lambda^\dag f^\dag\eta^{\dag 2} + \lambda f\eta^2)/2
 - g\eta^\dag d\eta\}, \label{AdS16}\\
V_{FI} & = & -\xi d, \label{AdS19}
\end{eqnarray}
where $V_{tree}$ is a tree level potential which is directly read from (\ref{AdS03}) and (\ref{AdS04}),
$V_{1-loop}$ is the dominant part of (\ref{AdS13}) with\footnote{The discrepancy between the definition of $\al$ here
and that in \cite{Ichi-Mura-PL2} is due to the assignment of the U(1) charge $g/2$ to$S_1$ and $S_2$ in the present
toy model.}
\begin{eqnarray}
\al \equiv \frac{l{\mit\Lambda}g^2}{128\pi^2}, \label{AdS17}
\end{eqnarray}
and $V_{FI}$ comes from the FI-term which has been activated in (\ref{AdS04}) owing to the choice of U(1) gauge
group. The tree potential (\ref{AdS15}) has a vanishing minimum at $f=d=\eta=0$ which is a necessary consequence of
SUSY model (\ref{AdS03},\ref{AdS04}). The FI-term (\ref{AdS19}) can spontaneously break SUSY when $d \neq 0$. As
is well-known, however, in case of $\la = 0$, i.e., no superpotential, $V_{tree} + V_{FI}$ has vanishing minima at
$|\eta| = \sqrt{-\xi/g}$ if $g\xi<0$ so that SUSY is not broken while the U(1) gauge symmetry is violated. In case of
$\la \neq 0$, these minima are raised to positive values so that SUSY is broken. In the following, we extend this fact
to the more general case of including $V_{1-loop}$ and show that it brings the minima back to zero making SUSY
restored for a specific value of $\al = 1$.   

$V_{eff}$ is a function of $f, d, \eta$ and $\al$ for given
$\lambda, g$ and $\xi$. We regard it an effective radion potential. As $\al \propto l = \pi \langle r\rangle$, it
does depend on the radion in spite that the radion does not directly couple with the fields on the "Planck brane".  

Hereafter we assume that VEVs are real for simplicity. The potential is minimized along the direction
$\pl V_{eff}/\pl f =
\pl V_{eff}/\pl d = 0$, i.e., for $f = \lambda(1-\al)\eta^2/2$ and $\ d = - \xi - (1+\al)g\eta^2$, which reduce the
potential to the function of two variables $\eta$ and $\al: \ V_{eff}(\eta, \al)$. 

If $g\xi > 0, V_{eff}(\eta, \al)$ has a minimum at $\eta = 0 \ (f=0, d=-\xi)$ for any $\al$ with a minimum value
$\xi^2/2$ which measures the SUSY breaking scale $M_{SUSY}$. $\al$ represents a flat direction and the radius of the
extra dimension is not stabilized. 

If $g\xi < 0$, on the other hand, $V_{eff}(\eta, \al)$ is shown in Fig.2 for tentative values $|\la| =|g| = 0.1$ and $|\xi| =
100$ TeV$^2$ and minimized for $\eta = \pm\sqrt{\displaystyle{ |\xi/2g} |}$ and $\al = 1 \ (f=d=0)$ with a vanishing
minimum value. Fig.3 shows the profile of $V_{eff}(\eta, \al)$ along $\pl V_{eff}/\pl \eta = 0$ i.e., for $\eta^2 =
-2(1+\al)g\xi/\{\la^2(1-\al)^2+2(1+\al)^2g^2\} \equiv {\tilde\eta}^2$;
\begin{eqnarray}
V_{eff}(\eta=\pm{\tilde \eta}, \al) =\frac{(1-\al)^2\lambda^2}{(1-\al)^2\lambda^2 +2(1+\al)^2g^2}\frac{\xi^2}{2},
\end{eqnarray} 
while Fig.4 shows the profile for $\al = 1$;
\begin{eqnarray}
V_{eff}(\eta, \al = 1) = 2\left(g\eta^2 + \frac{\xi}{2}\right)^2.
\end{eqnarray} 
Therefore, the radius of the extra dimension is stabilized at $\langle r\rangle = l/\pi = 32\pi/{\mit\Lambda}g^2$ making
SUSY restored in the true vacuum in spite of, or owing to, the presence of FI-term. The SUSY restoration is
realized for any $\xi$ that satisfies $g\xi < 0$. This result is extremely robust since taking the term $\propto \log
({\mit\Lambda}/ck)$ and the finite terms in (\ref{AdS13}) into account amounts only to shifting $\al$, hence $l$,
slightly.

It is easily confirmed that, in our mechanism of radius stabilization, the superpotential (\ref{AdS005}) plays an
essential role. If we turn off the superpotential by making $\la = 0, V_{eff}(\eta, \al)$ has a minimum at $\eta =
\pm\sqrt{\displaystyle{ |\xi/(1+\al)g} |}$ for any $\al$ with a vanishing minimum value, so that $\al$ represents a flat
direction and the radius of the extra dimension is not stabilized as in case of $g\xi > 0$. It is the contributions of the
auxiliary fields $F$ and $F^\dag$ that remove the flatness of $\al$ direction. Indeed, if $\la \neq 0$, we have
contributions of $F$ and $F^\dag$-tadpoles (Fig.1(a)) with the opposite sign to the contribution of $D$-tadpole (Fig.1(b))
up to the couplings $\la$ and $g$ so that they radiatively generate the nontrivial $\al$-dependence
of $V_{eff}(\eta=\pm{\tilde \eta}, \al)$ due to the bulk quantum effects.\footnote{The radius stabilization due to
quantum effects has been discussed in \cite{Gold-Roth} in a different context from us.} 

The restoration of SUSY with the vacuum energy being zero at the potential minima justifies our assumption that there
is no backreaction of fields in the bulk-boundary system to the background metric. 
\begin{figure}
\begin{center}
\includegraphics[width=100.5mm,height=88.55mm]{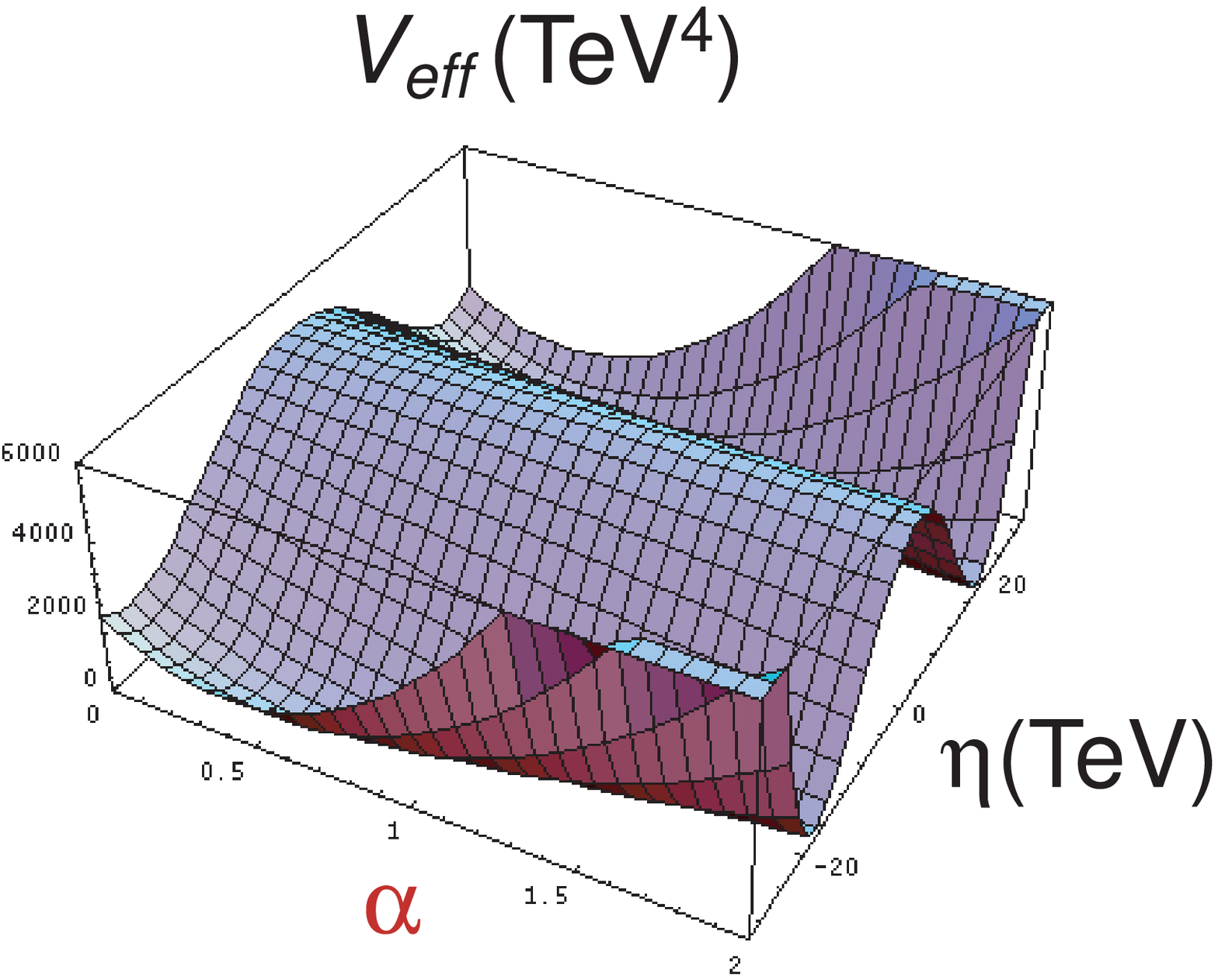}
%\BoxedEPSF{AdS5zu2.eps scaled 500}
%\centerpicture 130.5mm by 88.55mm ( Zu001 scaled 1000)
\end{center}
\caption{$V_{eff}(\eta, \al)$ for $|\la| = |g| = 0.1$ and $|\xi|=100$ TeV$^2$.}
\label{Fig2}
\end{figure}
\begin{figure}
\begin{center}
\includegraphics[width=100.5mm,height=88.55mm]{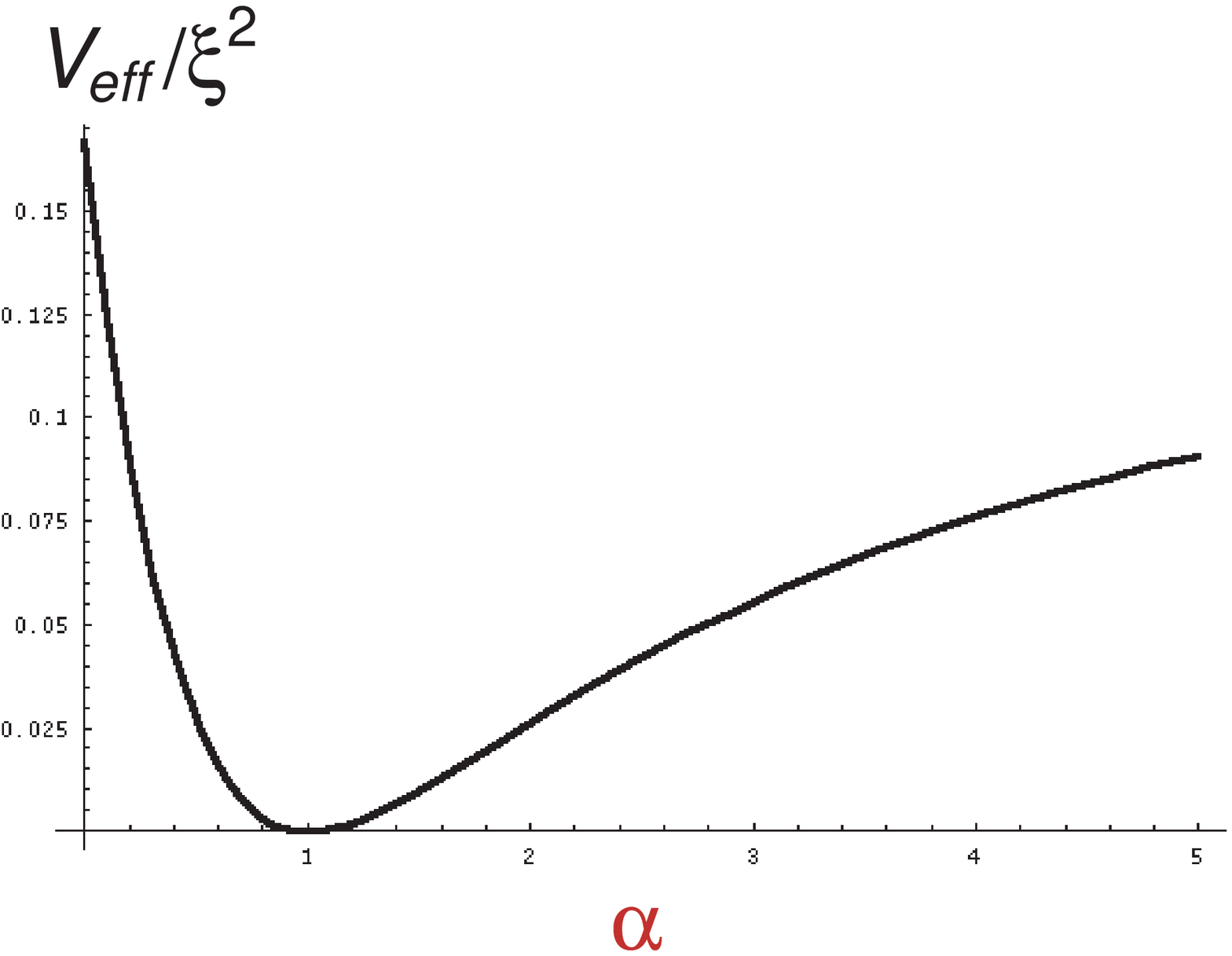}
%\BoxedEPSF{FIG04.eps scaled 500}
%\centerpicture 130.5mm by 88.55mm ( Zu001 scaled 1000)
\end{center}
\caption{Profile of $V_{eff}(\eta, \al)/\xi^2$ for $\pl V_{eff}/\pl \eta = 0$ and $|\la| = |g| = 0.1$.}
\label{Fig3}
\end{figure}
\begin{figure}
\begin{center}
\includegraphics[width=100.5mm,height=88.55mm]{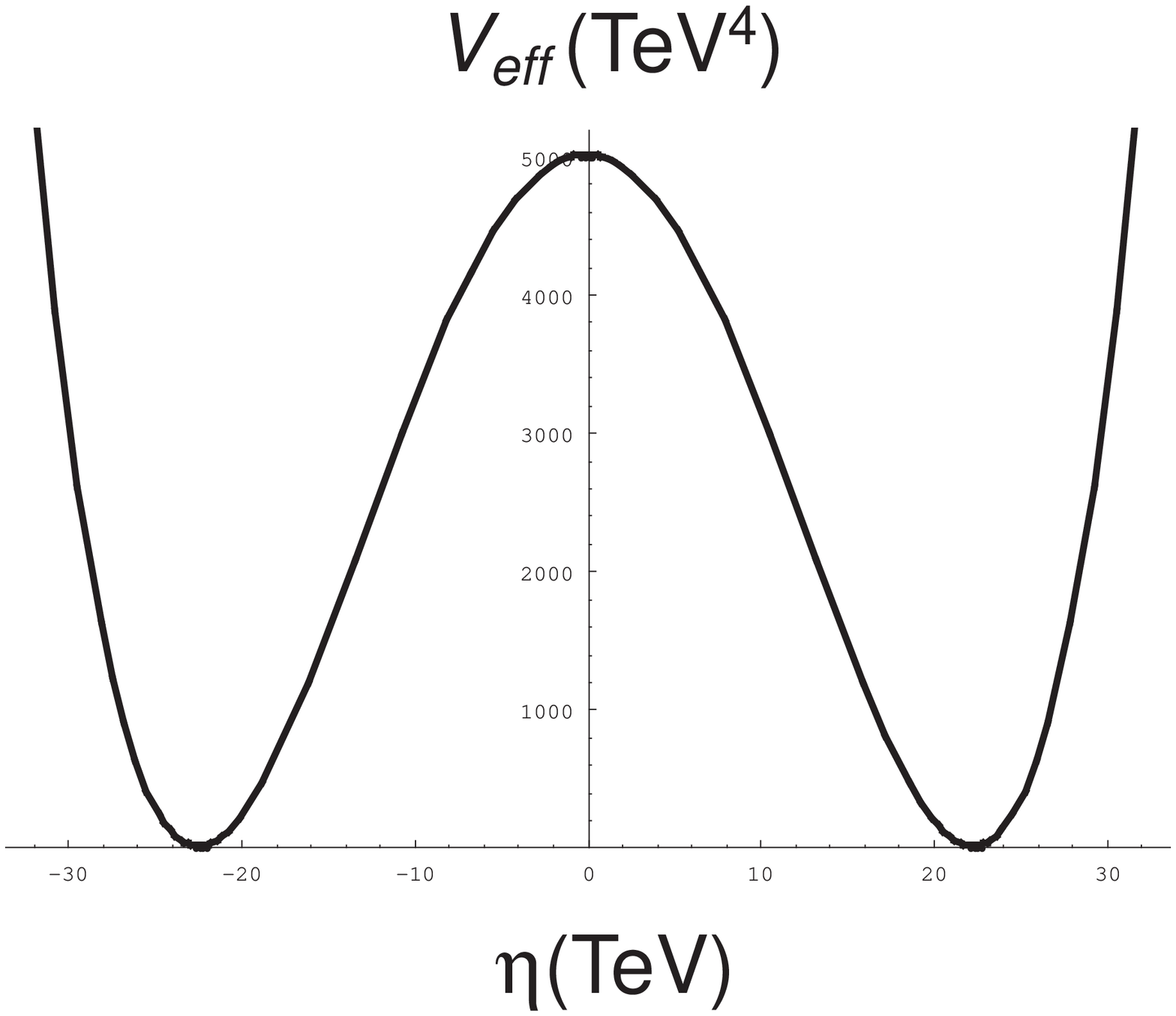}
%\BoxedEPSF{FIG05.eps scaled 500}
%\centerpicture 130.5mm by 88.55mm ( Zu001 scaled 1000)
\end{center}
\caption{Profile of $V_{eff}(\eta, \al)$ for $\al = 1, |g| = 0.1$ and $|\xi|=100$ TeV$^2$.}
\label{Fig3}
\end{figure}

One of interesting features of the result is the fact that this size of extra dimension corresponds to an intermediate
mass scale. For example, $l^{-1} \approx10^{11}$ GeV ($\approx 2 \times 10^{13}$ GeV) for
${\mit\Lambda} = M_{GUT} \approx 10^{16}$ GeV ($= M_{Pl}\approx 2 \times 10^{18}$ GeV) and $g\approx 0.1$ as
shown in Table \ref{tab1}. Thus, our $l$ is apparently incompatible with the assumption $kl \approx 35$ with $k
\approx M_{Pl}$. The value of $M_5$ is also shown in Table \ref{tab1} for two different cutoffs.
\begin{table}
\caption{The size of the extra dimension, 5D Planck mass and the radion mass estimated for $|\la| = |g| = 0.1, \al = 1$
and two different cutoffs.}
\label{tab1}
\begin{center}
\begin{tabular}
{|l|r|r|}\hline
${\mit\Lambda}$ & $M_{GUT}\approx 10^{16}$ GeV & $M_{Pl}\approx 2\times 10^{18}$ GeV \\
\hline\hline
$l^{-1}$              & $\approx 10^{11}$ GeV              & $\approx 2 \times 10^{13}$ GeV  \\
\hline
$M_5$ & $\approx 2 \times 10^{16}$ GeV & $\approx 2 \times 10^{17}$ GeV \\
\hline
$\langle r \rangle$ & $\approx 8 \times 10^{-13}$ fm & $\approx 3 \times 10^{-15}$ fm\\ 
\hline 
$m_{radion}$ & $\approx 0.03 (|\xi|/(100\mbox{TeV})^2)\mbox{GeV}$ & 
$\approx 10 (|\xi|/(100\mbox{TeV})^2)\mbox{GeV}$\\ 
\hline 
\end{tabular}
\end{center}
\end{table}

The radion mass is estimated for $g\xi < 0$ as
\begin{eqnarray}
m^2_{radion} \sim \left\vert\frac{\pl^2 f(r)}{\pl r^2}\right\vert^{-1}\left.\frac{\pl^2 V_{eff}}{\pl
r^2}\right\vert_{\al=1} \sim \frac{\la^2\xi^2 e^{2kl}}{k^2l^2M_{Pl}^2}, \label{AdS18}
\end{eqnarray}
where the canonical normalization of the radion kinetic term has been adopted. The radion mass strongly depends on the
parameter of FI-term $\xi$ which cannot be determined in the present model. 
\\
\\

\noindent{{\bf 5}\q {\bf Modified scales in Randall-Sundrum model}}\\

According to the conventional RS setup, $k\approx M_{Pl}$ and $q=$O(TeV) (hence the names {\it Planck brane} and {\it
TeV brane}) so that $l^{-1}$ is slightly smaller than $M_{Pl}$, i.e., $kl \approx 35$. It should be noted, however, that
the values of these parameters are not necessarily immovable. All they have to be subject to are constraints
(\ref{AdS00}), (\ref{AdS22}) and $k^{-1}\lesssim$ O(0.1mm). The constraint (\ref{AdS00}) tells us that
$q=$O(TeV) can be realized for $kl \approx 21 (\approx 27)$ when $l^{-1} \approx 10^{11}$ GeV ($\approx
2\times10^{13}$ GeV). Therefore, even if $l^{-1}$ is fixed at the intermediate scale, we can obtain the "TeV brane" by
setting $k$ at a scale an order or so greater than $l^{-1}$. 
%The 5D Planck scale $M_5$ is determined by (\ref{AdS22}) to be around a GUT scale $\approx 10^{16}$ GeV almost
%independently on the value of $l$. 

Thus, the radius stabilization suggested in \cite{Ichi-Mura-PL2} for the flat extra dimension can also be realized for the
5D super-YM model of the gauge group U(1) with warped extra dimension at the price of shifting the "Planck brane" to,
say,  an "intermediate brane". It is one of interesting problems to be addressed if the matter supermultiplets on this
brane could be one of ordinary SM particles. 

The numerical value of the radion mass (\ref{AdS18}) is shown in Table \ref{tab1}, where we have taken $\la = 0.1$ 
as a tentative value and $|\xi| = 10^4$ TeV$^2$ as a value of reference. This result suggests that $|\xi|
\gg 10^4$ TeV$^2$ or $|\xi| \ll 10^4$ TeV$^2$ might be required in order to avoid the cosmologically dangerous region.
\\ 

\noindent{{\bf 6}\q {\bf Conclusion}}\\

We have estimated the dominant 1-loop contribution to the 4D effective potential for a 5D super-YM model in the
warped background of AdS$_5$. If the matter supermultiplets are confined on the "Planck brane", it has been proved to be
proportional to the 4D momentum cutoff times the size of extra dimension, i.e., $V_{1-loop} \propto l{\mit\Lambda}$,
as in case of flat background. Choosing U(1) as the gauge group, we have minimized the effective potential and found
the case that the radius stabilization with $l^{-1} \approx 10^{11-13}$GeV can be realized, although the concept of {\it
Planck brane} should be modified to an {\it intermediate brane}.

Although the radion does not directly couple with the fields on the {\it intermediate brane}, our effective potential
can be regarded as a radion potential because the 1-loop contribution $V_{1-loop}$ is obtained by reducing the 5D
tadpole amplitudes to 4D and is proportional to $l = \pi\langle r\rangle$. 

Our model implements a new mechanism for the radius stabilization of the extra dimension in the following sense: (1) We
have a gauge supermultiplet in the bulk and chiral supermultiplets on the brane but need not introduce any hypermultiplet
in the bulk. (2) Interactions among the chiral fields on the brane through the superpotential plays an essential role in 
producing the nontrivial $l$-dependence of the effective potential due to the bulk quantum effects. 

It is a subject of future study to extend our model to the full 5D SUGRA taking the backreaction into account.\\

\noindent{{\bf Acknowledgment} \\

The authors would like to thank the participants in the Chubu summer school 2004 which was supported by
the Yukawa Institute for Theoretical Physics for useful discussions. 

%\newpage

\end{document}